# AI LITERACY OVER TOOL DESIGN: A MIXED-METHODS STUDY OF SCAFFOLDED VERSUS UNRESTRICTED GENERATIVE AI IN PROGRAMMING EDUCATION

**S. Azimi**
Delft University of Technology (NETHERLANDS)

## Abstract

Generative AI has become a routine resource in programming education, and most institutional responses to it are attempts at control, either by restricting access or by offering students a controlled version of the technology. This paper reports a seven-week mixed-methods pilot study conducted in a master's-level data analytics course, in which 33 students were randomly assigned either to a scaffolded AI Study Coach embedded in the notebook-based laboratory sessions or to unrestricted use of AI tools of their own choosing. The Coach offered stepwise hints in two modes, did not generate code, limited the number of hints available per session, and required a short reflection at the end of each session. The design assumed, in line with scaffolding theory and recent experimental evidence, that guided and limited support would build confidence and reduce over-reliance, and that the scaffolded group would learn more as a result. Assignment performance did not differ between the two conditions. Students in the Coach condition reported higher confidence but managed the hint budget poorly, while students in the unrestricted condition were satisfied with their tools and, at the same time, uneasy about the degree to which they depended on them. In interviews, students in both conditions identified awareness of their own reliance on AI as the most valuable outcome of the course. Students who had formulated their own rules for when to use AI performed better in both conditions, and those with the best understanding of how the models work, which was in every case self-taught, used the tools most deliberately and achieved the highest scores. The design of the tool proved to matter less than the students' capacity to govern their own use of it, a capacity that is at present acquired by chance. The paper argues that the appropriate response is structural: assessment that grades the reasoning behind AI-assisted work, and AI literacy taught explicitly as a core skill.

## 1 INTRODUCTION

Generative AI is present in every programming course, whether or not the course permits it. The Digital Education Council's 2026 global survey, which collected more than 45,000 responses from students and faculty in 35 countries, found that 88 percent of students use AI in their learning and that 57 percent consider the guidance they receive on its use in assessment to be inadequate [1]. Institutions have responded in two ways. The first response is restriction, in the form of bans, disclosure requirements and automated detection. The evidence on detection is not encouraging: in a systematic test of fourteen detection tools, Weber-Wulff et al. found that none reached 80 percent accuracy, and OpenAI withdrew its own classifier in the same year on the grounds that its accuracy was too low [2]. The second response is tool design, in which institutions build or license tutors, hint generators and adaptive systems that give students a controlled version of the technology. Yan et al. reviewed 118 studies of large language models in education and classified them into 53 use cases across nine categories of educational task, from grading and feedback to content generation and recommendation [3]. All nine categories concern tasks performed for or on the student; none concerns the student's own regulation of how the tool is used.

Programming courses present a further difficulty that AI is often presented as solving. Students enter such courses with widely varying levels of programming experience, and the gap between them is largest among the least confident students, who are also the students most likely to use a generative tool as a substitute for their own work. A tool that adapts its support to the needs of each student appears to offer a way of closing that gap.

The pilot study reported in this paper set out to test that promise. Its design followed scaffolding theory, according to which novices learn more from structured support that is gradually withdrawn than from unrestricted access to help [4]. Whether scaffolded support leads to learning depends on the learner's self-regulation, that is, on how students plan, monitor and evaluate their own work [5]. Long and Magerko define AI literacy as a set of competencies that enables individuals to evaluate AI technologies critically and to use them effectively, and they argue that these competencies have to be taught [6]. The design assumption was also consistent with the strongest experimental evidence available at the time. In a field experiment with nearly a thousand secondary-school students, Bastani et al. found that unrestricted access to a GPT-4 tutor improved performance during practice but reduced performance on a subsequent examination taken without AI, whereas a version of the tutor with safeguards, which provided hints without giving answers, removed the harm [7].

On this basis the study assumed that a scaffolded coach would build confidence and reduce over-reliance, and that students working with it would learn more than students with unrestricted access to AI. Section 2 describes the setting, the Coach and the data collected. Section 3 reports the results, which did not confirm the assumption. Section 4 discusses the results in the light of the learning sciences and draws the implications for course design.

## 2 METHODOLOGY

### 2.1 Setting and participants

The pilot was conducted over seven weeks in a master's-level course. The course teaches data analysis and machine learning through weekly programming laboratory sessions in Jupyter notebooks, and its students come from several master's programmes with different amounts of prior programming experience. Thirty-three students took part in the study. They were randomly assigned to one of two conditions. Students in the intervention condition (n=16) worked with an AI Study Coach that was built into the laboratory notebooks and did not use other AI tools during the sessions. Students in the comparison condition (n=17) were free to use any AI tools they chose, in whatever way they chose. The teaching team consisted of the course lead and three teaching assistants, two of whom were master's students and one a doctoral candidate. Participation in the data collection was voluntary, and all data were pseudonymised before analysis.

### 2.2 The AI Study Coach

The Coach was integrated into the notebook environment so that students could consult it without leaving the laboratory task. It operated in two modes. In Explainer mode it explained analytic outputs and the design choices behind the code in the notebook. In Debugger mode it responded to a student who was stuck with a sequence of hints that moved from a general pointer towards a more specific one, and it did not provide complete solutions or generate code. Four constraints applied in every session: a budget of 25 hints per session, a maximum of fifteen minutes of AI use per session, a short written reflection at the end of each session, and a requirement that every piece of AI-sourced code be tagged as such in the notebook. Together, these constraints constituted the scaffold. The budget was intended to make each request deliberate, the time limit to keep the student's own work at the centre of the session, the reflection to prompt monitoring, and the tagging to keep the use of AI visible to the teaching team.

### 2.3 Data collection

Five sources of data were collected in both conditions. A concept inventory consisting of ten scenario items on fairness, data leakage and model evaluation was administered in week 2 and again in week 7. Assignments were scored by the teaching assistants with a rubric that covered code correctness, error diagnosis, reasoning and the responsible use of AI. Surveys administered before and after the series of sessions asked students about their confidence, their reliance on AI, their critical evaluation of AI output and their conceptual understanding. The Coach logged every interaction, including the number of hints requested, the mode used, the timing of requests and the point in the task at which help was sought. Six semi-structured interviews were conducted with students from both conditions, and informal discussions with students and teaching assistants during the sessions were recorded in notes.

### 2.4 Analysis

Gains on the concept inventory and scores on the assignments were compared descriptively between the two conditions. With sixteen and seventeen students per condition, the pilot was not designed to support inferential tests, and none are reported. The interaction logs were analysed for patterns of use, with particular attention to how students distributed their hint budget over a session. The interviews and the end-of-session reflections were coded thematically. The three strands of data were then read together, with the logs and interviews used to interpret the comparisons of survey responses and scores.

## 3 RESULTS

### 3.1 Performance and confidence

The two conditions did not differ in assignment performance. Gains on the concept inventory between week 2 and week 7 were similar in both groups, and the rubric scores on the assignments were equivalent. The scaffold did not produce the learning advantage that the design had assumed.

Confidence, however, did differ. In the post-session surveys, students in the Coach condition reported higher confidence in their own competence than students in the unrestricted condition, and the gain in reported confidence over the seven weeks was larger in the Coach condition. Students in both conditions reported that AI had helped them to understand analytic output, to debug code and to clarify unfamiliar concepts, and students in the Coach condition rated the Coach's explanations positively for the clarification of concepts.

### 3.2 How students used the hint budget

The hint budget did not function as the rationing mechanism it had been designed to be. The interaction logs show two distinct patterns of use. Some students spent most of their budget early in a session, on the first problems they encountered, and had few hints left for the more demanding tasks at the end of the session. Others saved their hints for later and finished the session with most of the budget unused. Several students complained about the limit in their reflections and in the interviews, and a few reported that they had looked at the screens of classmates in the unrestricted condition when they were stuck.

Within the Coach condition, the students who had developed a deliberate strategy for when to spend a hint and when to hold back scored higher on the assignments than the students who saved their hints without a plan. The budget rewarded a capacity for self-governance that some students already possessed; it did not develop that capacity in the students who lacked it.

### 3.3 Awareness, rules and understanding

The finding that students valued most was not part of the design. In the interviews, students in both conditions said that the most useful outcome of the laboratory sessions had been becoming aware of how much they relied on AI. Students in the unrestricted condition were satisfied with their tools and, in the same interviews, uneasy about that reliance. They described their engagement with the material as shallow, said that they felt their analytical skills slipping, and said that they nevertheless could not stop, because the tools were too efficient to give up. Several of them said, without being asked, that there ought to be a way of using AI that gave them its full capacity while still keeping them intellectually engaged. Students in the Coach condition reported that the budget had followed them out of the classroom: when working with unrestricted tools at home, they found themselves asking whether a question was worth a hint before putting it to the tool. One student described the comparison in these terms: "I go through it myself, redoing it. Most of the time what I do is better. There's more depth when you do it yourself."

Awareness on its own did not change how students worked. The students whose practice did differ were those who had set themselves a rule. Several students in both conditions had formulated their own rules for the use of AI, either before the course or during it: building a solution first and using AI only to check it, delegating syntax to the tool while writing the logic themselves, or withholding part of the problem from the tool in order to force their own reasoning. These students performed better on the assignments regardless of the condition to which they had been assigned.

Students' understanding of how the language models work varied widely across the cohort, and in every case it was self-taught, since the course did not teach it. The students with the best understanding of the models wrote more effective prompts and used the tools more deliberately, and they also achieved the highest assignment scores. Before generative AI, the advantage in a programming course went to the student who could code most fluently. In this cohort it went to the student who governed their use of AI most effectively, and that advantage was distributed by prior exposure and by disposition, not by anything the course had done.

## 4 CONCLUSIONS

The pilot tested an assumption that is common in current practice: that a scaffolded and controlled version of generative AI will serve students better than open access to it. On assignment performance the two conditions were indistinguishable. The scaffold raised confidence and made students more aware of their reliance on AI, but it did not produce either of the two things that separated the strongest students from the rest. Those students had rules for when to use AI, and they understood what the tools were doing. The course had provided neither.

Three readings from the learning sciences are consistent with the results. First, scaffolding reduced the cognitive load of individual tasks and raised confidence, which is what it is designed to do [4], [8]; it did not build the capacity to govern the tool, which is a different outcome. Second, self-determination theory predicts that self-regulation develops under conditions of autonomy [9]. The students who complained about the limit and the students who set their own rules point in the same direction: the restriction did not produce self-governance, and the students who governed themselves had chosen to do so. Third, governing one's own use of AI is a self-regulatory skill in Zimmerman's sense [5], and like other such skills it develops through monitoring, reflection and visible process, not through restriction alone. Fan et al. report a comparable pattern in a larger randomised study: access to ChatGPT raised the quality of students' essays without improving their knowledge gain or transfer, a result the authors describe as metacognitive laziness [10]. In the present study, too, equal scores concealed unequal understanding.

Two limitations bound these conclusions. The pilot was small and was conducted in a single course, so the direction of the findings is clear but their magnitude is not. The comparison was between one particular scaffold and open access, and the results say nothing about scaffolds in general, only that this reasonable one did not do what it was assumed to do.

The implication for course design is structural. Generative AI develops quickly, and a course or tool built around one version of it is out of date by the time it runs; educational tools cannot keep pace with commercial systems, which now offer study modes of their own. Assessment, on the other hand, can be designed to target the reasoning behind AI-assisted work and to require students to reason before they consult AI, so that how a student worked becomes visible and gradable. The rules that the strongest students set for themselves can be taught, and students can be asked to write, test and revise rules of their own. AI literacy in the sense of Long and Magerko, including a working understanding of what the models do, belongs in the curriculum for students and for staff alike. A framework for course redesign along these lines is developed in [11].